\documentclass{article}
\usepackage[utf8]{inputenc}
\usepackage{graphicx}
\usepackage{pdfpages}
\usepackage{dcolumn}
\usepackage{bm}
\usepackage{color}
\usepackage{sidecap}
\usepackage{array}
\usepackage{amsmath}
\usepackage{multirow}
\usepackage{float}
\usepackage{hyperref}
\usepackage{geometry}
\usepackage{siunitx}
\usepackage{subfig}
\title{\bf Influence of ionic conditions on knotting in a coarse-grained model for DNA}
\author{S. Wettermann, R. Datta, P. Virnau}
\date{}

\begin{document}
\maketitle
\begin{abstract}
We investigate knotting probabilities of long double-stranded DNA strands in a coarse-grained Kratky-Porod model for DNA with Monte Carlo simulations. Various ionic conditions are implemented by adjusting the effective diameter of monomers. We find that the occurrence of knots in DNA can be reinforced considerably by high salt conditions and confinement between plates. Likewise, knots can almost be dissolved completely in a low salt scenario. Comparisons with recent experiments confirm that the coarse-grained model is able to capture and quantitatively predict topological features of DNA and can be used for guiding future experiments on DNA knots.
\end{abstract}

\section*{Introduction}
The revelation of DNA packing and folding in the cell nucleus \cite{Lieberman-Aiden_2009,Stevens_2017,Siebert_2017,Ganji_2018} and the emergence of commercially available nanopore techniques \cite{Jain_2016,Jain_2018} has ushered in a new era of DNA research in the past decade. Knots, which emerge naturally as a byproduct in long macromolecules like DNA \cite{Frisch:JACS:1961,Delbruck_knot_62}, may however be detrimental to biological processes and technical applications. It is therefore of prime importance to study conditions and length scales at which they appear in equilibrium and develop strategies to enhance \cite{Lua_2004,Virnau:JACS:2005,Tang:PNAS:2011,Amin_2018} or suppress knotting \cite{renner:ACSmacroL:2014,diStefano:Softmatter:2014}. From a technical point of view, numerical simulations are a great tool for this task as structural and topological information are readily available. Coarse-grained models are particularly relevant as knots appear at scales beyond the Kuhn length and models with atomistic resolution are often poorly suited for efficient Monte Carlo algorithms required to scan configuration space. It is therefore crucial to test and improve coarse-grained models for DNA to quantitatively support and interpret experimental efforts.

A first link to double-stranded (ds) DNA was already established in the first simulation paper on polymer knots from 1974 \cite{Vologodskii:1974}. In their seminal contribution, Vologodskii et al. determined knotting probabilities of random walks and associated single segments with the Kuhn length of DNA ($100~nm$) - a prediction which turns out to be surprisingly accurate as we will demonstrate later. This basic approach has been refined further in the early 1990s in conjunction with gel electrophoresis experiments on short DNA strands of up to 10~kbp \cite{Rybenkov:1993:PNAS, Shaw1994}. Ideal segments were replaced by cylinders with excluded volume interactions that depend on ionic conditions \cite{Rybenkov:1993:PNAS}, and it was also demonstrated that DNA knotting probabilities vary somewhat with solvent conditions (reaching about $4\%$ in a high salt environment.) Higher resolution versions of this model in which one Kuhn length is represented by several segments have been used to study the effect of confinement on short strands in high salt conditions. 
Among other things, Orlandini, Micheletti and coworkers have demonstrated with numerical simulations that confining DNA between plates or in nanopores increases knotting probabilities when typical distances between plates or nanopore diameters are in the order of the Kuhn length of DNA \cite{Micheletti:Macromol:2012,Micheletti:SoftMatt:2012,Orlandini2013}. Alternatively, coarse-grained bead-stick \cite{Dai_2012_ACS,Dai_2012_SM,Rieger:PLoS:2016} or bead-spring \cite{Trefz:PNAS:2014,Rothoerl2022} representations for DNA can be used in which the effective diameter is adjusted to account for varying solvent conditions and which can be adapted for molecular dynamics simulations. Of particular relevance to our study is  \cite{Dai_2012_ACS} in which the authors have studied knotting of closed DNA rings in bulk and plate geometry and to which our results for open strands can be compared. Variants of this model class have also been applied to investigate, e.g., statics \cite{Dai_2015,Jain_2017} and dynamics of DNA knots in a nanochannel \cite{Micheletti:MacroLett:2014}, packing of DNA in viral capsids \cite{MarenduzzoPNAS2009,Reith:NAR:2012} and recently for the reproduction of experimental knotting probabilities of $\lambda$ phage DNA in high salt conditions \cite{KumarSharma2019}. Of course, there are also limits to this class of coarse-grained descriptions, and higher resolution models \cite{Suma_2017,Suma_2018} may address questions which either require a detailed structural description or an explicit modelling of electrostatic interactions \cite{Suma_2018}.

In this work we systematically extend previous analyses to DNA lengths relevant to modern experiments on $\lambda$ \cite{Plesa2016,KumarSharma2019} and $T4$ phages \cite{Plesa2016}. Our comprehensive study also covers the full range of ionic conditions for free DNA and DNA confined between two plates, and comparisons with existing experimental data confirm the validity of the modelling approach.
This enables us to show, amongst others, that for the considered strand sizes the dependence of knotting on salt concentrations  \cite{Rybenkov:1993:PNAS} can be used to effectively disentangle DNA prior to experiments where knots are undesired.

\section*{Method}
\begin{figure}[htp]

	\centering
	\begin{minipage}[b]{0.6\textwidth}
		\includegraphics[width=\linewidth]{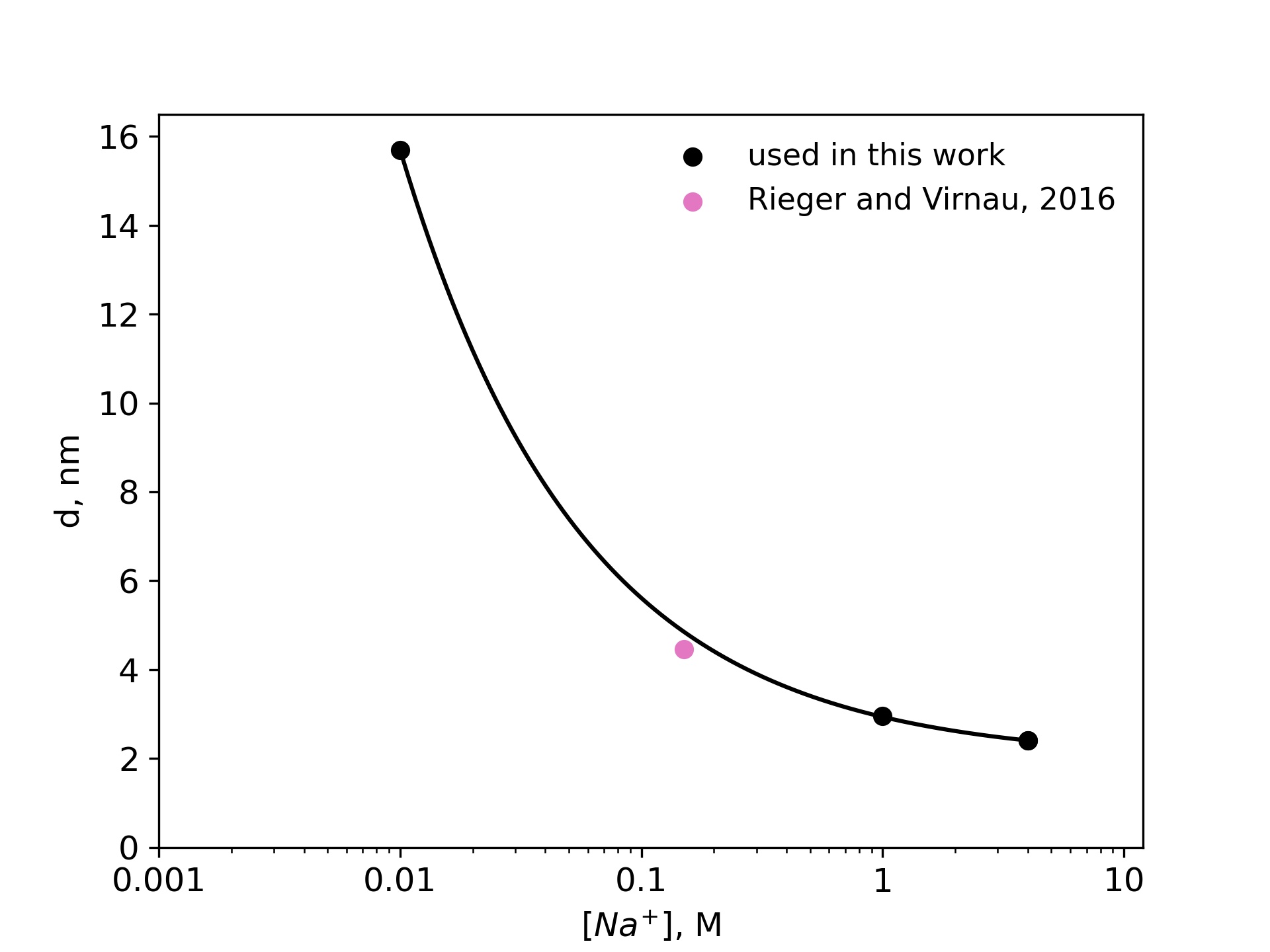}
		\label{fig:stigter}
	\end{minipage}  
	\begin{minipage}[b]{0.61\textwidth}
		\includegraphics[width=\linewidth]{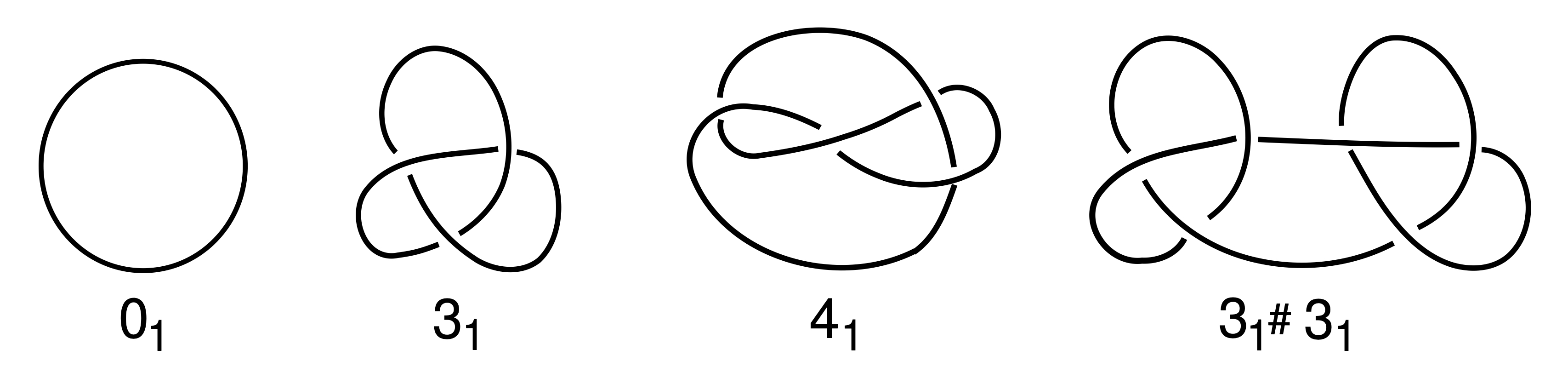}
		\label{fig:knot_types}
	\end{minipage}
	\caption{\textbf{(a)} Effective diameter of DNA, $d$ as a function of NaCl concentration as calculated by Stigter on the basis of polyelectrolyte theory \cite{Stigter1977}. The values for the effective diameter, $d$ have been obtained by calculating the interaction potentials of highly charged colloidal cylinders representing DNA. The effective diameter value for physiological salt concentrations $0.15$M has been taken from \cite{Rieger:PLoS:2016}. 
		\textbf{(b)} Knot types considered in this paper.}
	\label{fig:fig1}
\end{figure}

\textbf{Implicit modelling of ionic solvent conditions.}

DNA is negatively charged, but long-range electrostatic interactions are partially or completely screened by counterions in the solution. In this paper we follow an implicit solvent approach pioneered by \cite{Stigter1977} and \cite{Rybenkov:1993:PNAS}
in which screened charges are represented by effective excluded volume interactions.
The diameter $d$ of a DNA chain is a parameter that quantifies the latter and can be defined as the segment diameter of a representative chain which is uncharged, but has the same configurational and morphological properties as the original DNA with partial or completely screened charges. The magnitude of the electrostatic repulsion, and consequently, the numerical value of $d$, is a function of salt concentration. Stigter \cite{Stigter1977} modeled DNA in sodium chloride solution as charged cylinders. Following the theory developed by McMillian and Mayer \cite{McMillan_Meyer_1945} and the calculations of Hill \cite{Hill_1956,Hill_1960}, Stigter carried out analytic calculations to estimate the effective diameter of DNA as a function of sodium chloride concentration (see Fig.~\ref{fig:fig1}a). 

Already in 1993 Rybenkov et al. \cite{Rybenkov:1993:PNAS} were able to confirm this approach (and Fig.~\ref{fig:fig1}a) by representing DNA as a closed chain of cylinders of Kuhn length 100~nm and by matching experimentally determined knotting probabilities of a short $P4$~phage DNA strand (of around 10,000 base pairs) with those obtained from Monte Carlo simulations. 

Here, we use a higher resolution variation of this ansatz which models DNA as a standard bead-stick chain and also resolves local structure at the scale of the persistence length (which according to Kratky-Porod theory is half of the Kuhn length). We keep, however, the same effective diameter to determine knotting probabilities in various ionic conditions for long, experimentally relevant DNA strands (like $\lambda$ phage or T4). In a previous work \cite{Rieger:PLoS:2016}, we have already validated this approach by determining simulation parameters for physiological conditions ($0.15M$) that reproduce experimental knotting spectra of short strands from \cite{Rybenkov:1993:PNAS} and \cite{Shaw1994} even without making assumptions about the persistence or Kuhn length. Not only did these simulations confirm a value for $d$ which is close to the value of Stigter (pink point in Fig.~\ref{fig:fig1}a), they also confirmed the correct persistence length of DNA. While we use $d$=4.465~nm for physiological conditions, values for other ionic conditions are directly taken from Fig.~\ref{fig:fig1}a.

\noindent
\textbf{Bead-stick model.}
Simulations were performed using a discrete Kratky-Porod model \cite{KratkyPorod1949,Dai_2012_ACS,Rieger:PLoS:2016,Marenz_2016} with hard sphere interactions between monomers and a constant distance between adjacent beads. For simulations in slit confinement, walls are also hard and impenetrable.
Chain stiffness is implemented via a bond-bending potential:
\begin{equation}
	U = \kappa \sum_{i} (1-\cos\theta_{i})
\end{equation} where $\theta_{i}$ for $i = 1, \dots, N-1$ are the angles between adjacent bond vectors. Simulations were performed at various salt concentrations with values for $d$ obtained from Fig.~\ref{fig:fig1}a.
We assume a persistence length $l_p$ of 50 nm or 150 base pairs (bp) for all considered salt concentrations. For a Kratky-Porod chain the stiffness parameter $\kappa$ for any given effective diameter $d$ can be computed as \cite{Fisher1964,Trefz:PNAS:2014,Rieger:PLoS:2016}
\begin{equation}
	\kappa \approx \frac{l_{p}\cdot k_{B}T}{d}.
	\label{eq:stiffness}
\end{equation}
In dsDNA the distance between adjacent base pairs is $1/3$~nm. By comparing the contour lengths, we conclude that a DNA strand with $B$ base pairs is represented by a chain of 
\begin{equation}
	N\approx B\cdot 0.3333 ~\textrm{nm}/d    
	\label{eq:contour}
\end{equation}
beads.
\\
Several simplifications are implied in this approach. The dependence of persistence length on ionic conditions was neglected as differences only amount to a few percent at least in the formalism of Odijk \cite{Odijk_1977}, Skolnick and Fixman \cite{Skolnick_1977}.
Note, however, that for small DNA strands (up to several kilo bases) and high salt conditions the persistence length can be significantly smaller $(\approx30-35~nm)$, \cite{Savelyev_2012, Brunet2015, Kam_1981, Manning_1981, Post_1983, Rieger_2018} and also depends on the specific ions in the solvent \cite{Brunet2015} (the influence of which we neglect as well). Nevertheless, for larger chains (such as those simulated in our paper) persistence length is expected to increase again and might actually be closer to $50~nm$. In high salt conditions knotting probabilities also depend little on the actual value of the persistence length as demonstrated in Supplementary Information, which taken together justifies our simplified assumptions.

Experiments \cite{Plesa2016,KumarSharma2019} displayed in Fig.~\ref{fig:fig2}a use either KCl (1 and 1.5M) or LiCl (4M) as buffer.
In our simulations we mainly study DNA strands of lengths
$20,678$~bp, $48,502$~bp and $165,648$~bp corresponding to a linearized plasmid, $\lambda$~phage DNA and phage $T4~GT7$ DNA used in \cite{Plesa2016}.

For comparison we have also implemented a simple random walk which can be mapped onto DNA by setting the Kuhn length to 100~nm, which takes over the role of $d$ from Eq. \eqref{eq:contour}. Interestingly, this simplistic model for DNA was already discussed in the first simulation paper on polymer knots from 1974 \cite{Vologodskii:1974} and yields, as we will see later, surprisingly reasonable results when compared with recent experiments on long DNA strands \cite{Plesa2016}. Of course, differences in knotting probabilities due to varying ionic conditions are not captured in this approach, but could in principle be included following \cite{Rybenkov:1993:PNAS}.
All chains were simulated with a pivot Monte Carlo algorithm \cite{Madras1988}: After a pivot center is chosen at random, one arm of the polymer is rotated by a random angle around the pivot point and the move is accepted with the Metropolis criterion. 

\noindent
\textbf{Knot analysis.}Knots are defined only for closed chains and characterised by the minimum number of crossings when projected onto a two-dimensional plane (see Fig.~\ref{fig:fig1}b)\cite{Adams:1994}. The simplest knot, apart from an unknotted ring which is called the unknot ($0$), is the trefoil ($3_1$) with three essential crossings. Similarly, the next knot type to follow is the figure-eight knot ($4_1$) with four crossings. While there is only one knot with three and one knot with four crossings (as indicated by the index), eventually the number of different knots grows exponentially with the crossing number.
In addition to prime knots, multiple knots can also be combined on a ring to form so-called composite knots as indicated in the right-most picture of Fig.~\ref{fig:fig1}b. 

Since we have simulated linear chains a closure to connect the two end points of our chains needs to be defined. For this we first connect the two termini with their centre of mass. Along these lines one can then define a closure which emerges from one end, follows the first line, connects to the second one far away from the polymer and ends at the second end of the chain \cite{Virnau:PLoScb:2006}. Once the open chain has been closed the Alexander polynomial can be determined for which a detailed description can be found in \cite{Virnau:Phys.Proc:2010}. The size of a knot can be determined by successively removing monomers from the two ends of the polymer (before closure) chain until the knot type changes.

\section*{Results}
\begin{figure}[htp]
	\centering
	\begin{minipage}[b]{0.6\textwidth}
		\includegraphics[width=\linewidth]{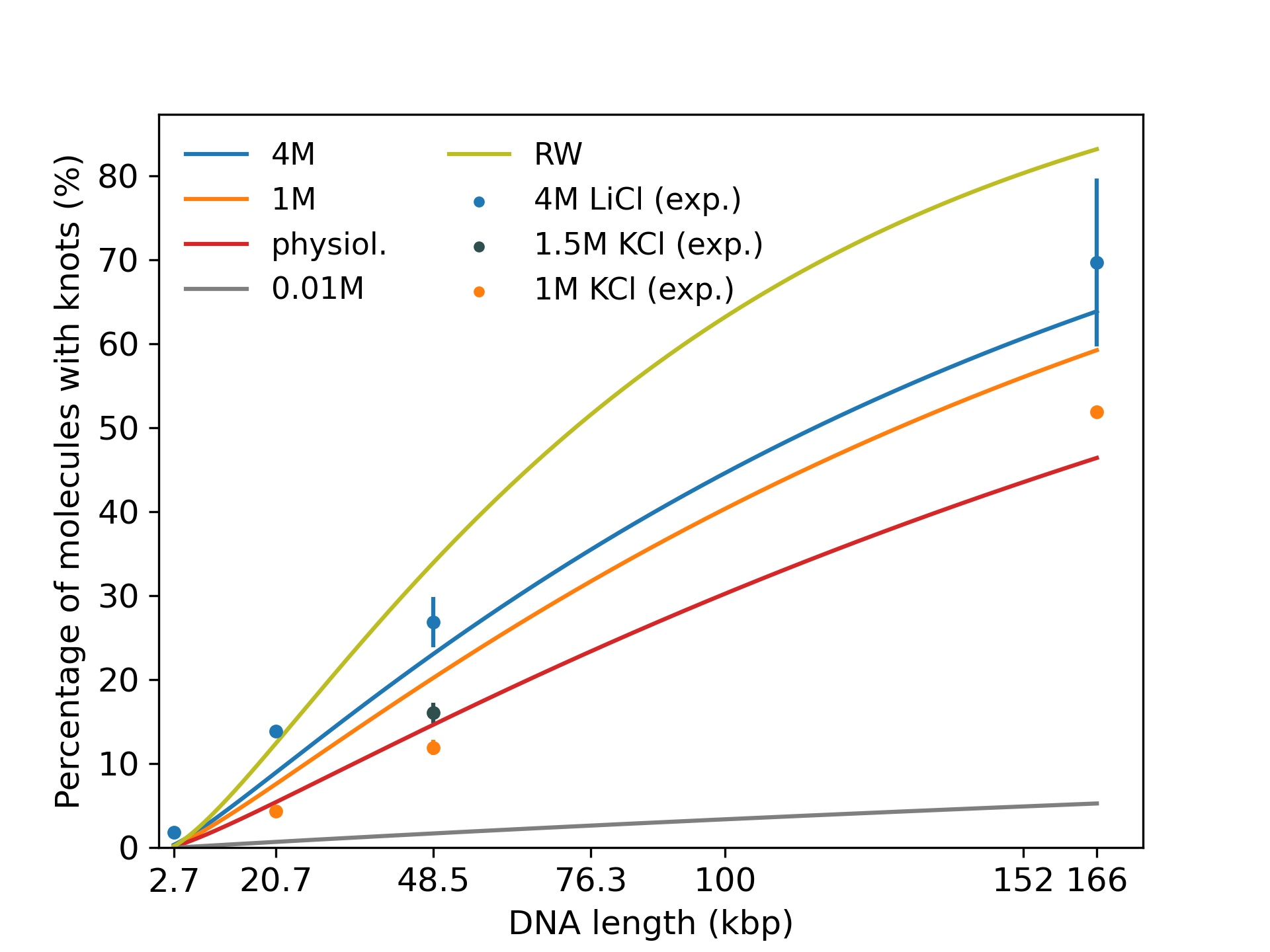}
		\label{fig:Subfigure2_1}
	\end{minipage}  
	
	\begin{minipage}[b]{0.6\textwidth}
		\includegraphics[width=\linewidth]{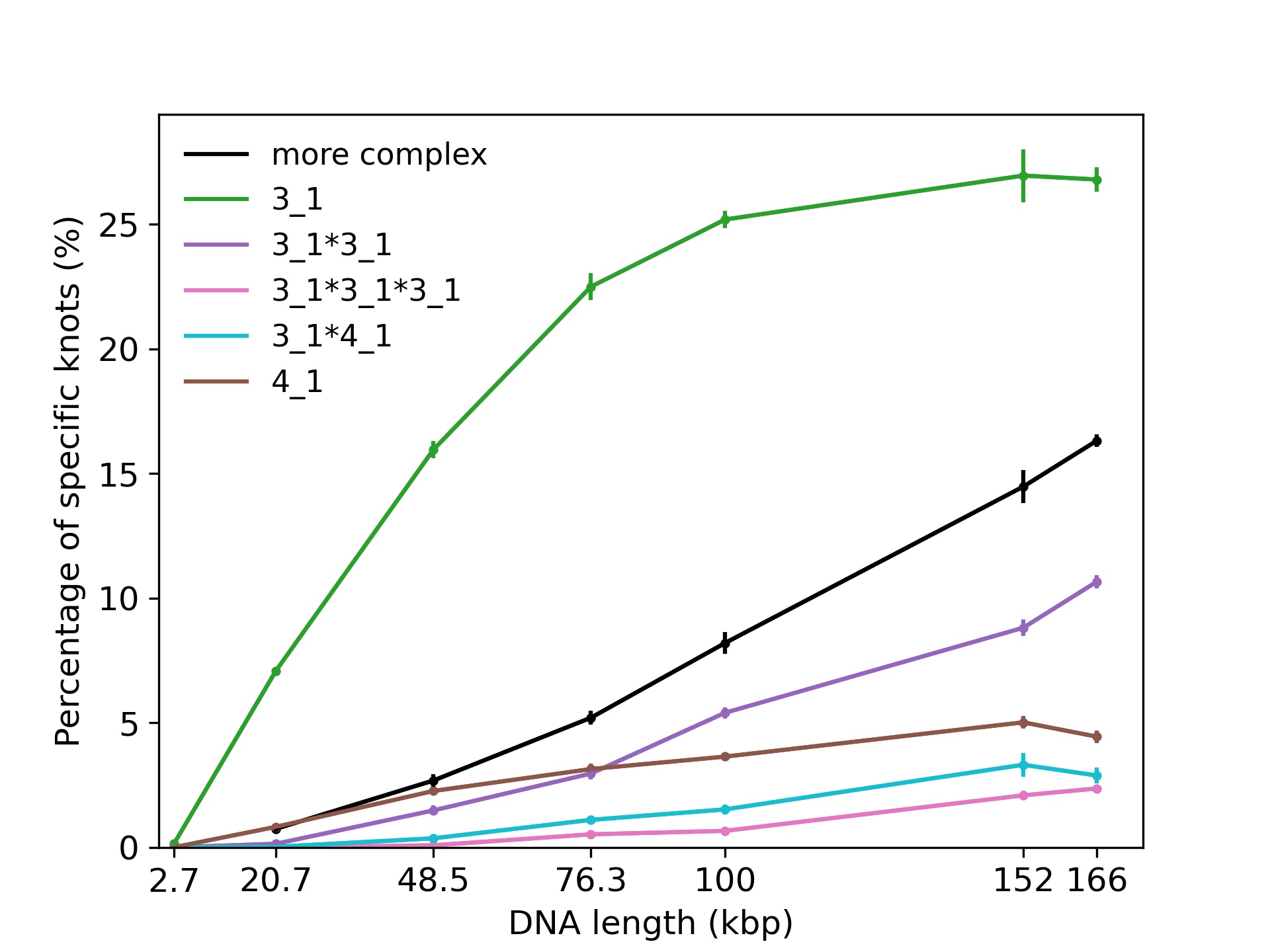}
		\label{fig:Subfigure2_2}
	\end{minipage}

	\caption{\textbf{(a)} Knotting probability as a function of chain length for DNA chains of up to 166~kbp in comparison to recent experiments \cite{Plesa2016,KumarSharma2019}. Straight lines refer to simulations and data points to experiments. Different colors represent varying ionic conditions. \textbf{(b)} Knot spectrum as a function of DNA length for simulations in a high salt solvent ($4$M) scenario. Complex knots refer to all other knots not listed here excluding the unknot.}
	\label{fig:fig2}
\end{figure}

First, we investigate knotting probabilities as a function of DNA length and ionic conditions for unconfined DNA (Fig.~\ref{fig:fig2}a). For better clarity, we only plot fitted curves according to \cite{Deguchi_1997} for our simulated data. Experimental data from recent nanopore experiments \cite{Plesa2016,KumarSharma2019} on 20,678 bp long linearized plasmids, $\lambda$ phage (48,502 bp) and $T4~GT7$ DNA (165,648 bp) are displayed as data points. We notice that even at the range of 100 kbp, DNA already exhibits substantial knotting which strongly depends on ionic conditions. Knotting probabilities are larger in high salt scenarios and can reach up to 70$\%$ for the largest strands. Intriguingly, our coarse-grained simulations also suggest that knotting can almost be avoided completely in a low salt scenario. For the same 166~kbp strand we only observe a knotting probability of 5$\%$, which (if confirmed experimentally) would open up new possibilities for disentangling large DNA strands, e.g., in preparation of nanopore sequencing. The latter may, however, prove challenging experimentally as it becomes difficult to translocate at low ionic concentrations. These large discrepancies are indeed surprising as prior simulations of closed DNA rings with a similar model yielded significantly higher knotting probabilities, particularly for the low salt scenario \cite{Dai_2012_ACS}.
Overall, agreement between predicted and experimentally determined knotting probabilities in medium to high salt conditions is quite good and differences only amount to a few percent. Surprisingly, comparisons with a simple random walk model still yield reasonable agreement even though occurrences of knots are overestimated systematically.
At the length scales considered, the knot spectrum is still dominated by trefoil knots as is depicted for the high salt (4M) scenario in Fig.~\ref{fig:fig2}b. However, we already observe the emergence of composite knots as demonstrated before for even larger chains under physiological conditions in \cite{Rieger:PLoS:2016}. 

\begin{figure}[htp]
	\centering
	\begin{minipage}[b]{0.6\textwidth}
		\includegraphics[width=\linewidth]{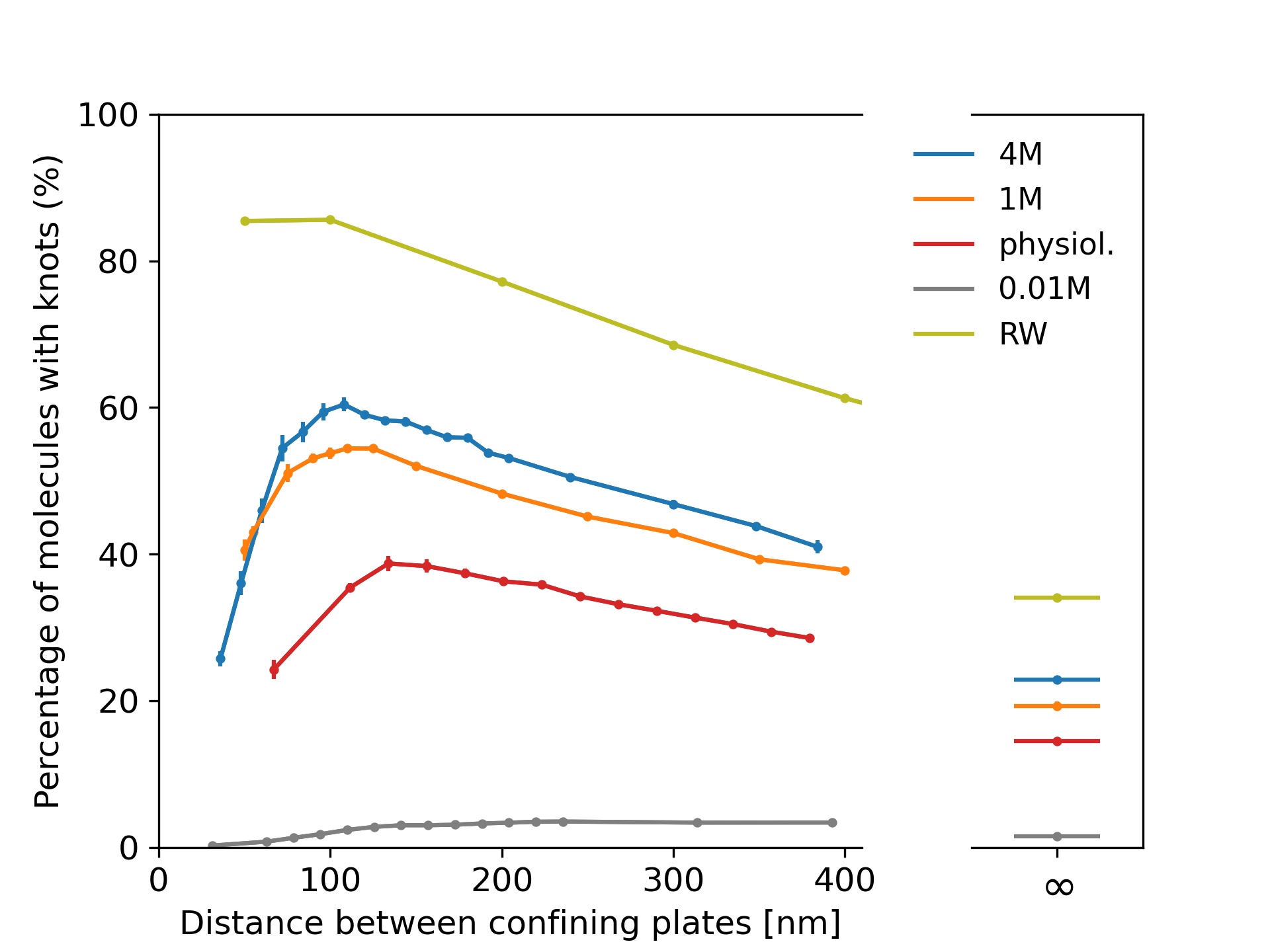}
		\label{fig:Subfigure3_1}
	\end{minipage}%
	
	\begin{minipage}[b]{0.6\textwidth}
		\includegraphics[width=\linewidth]{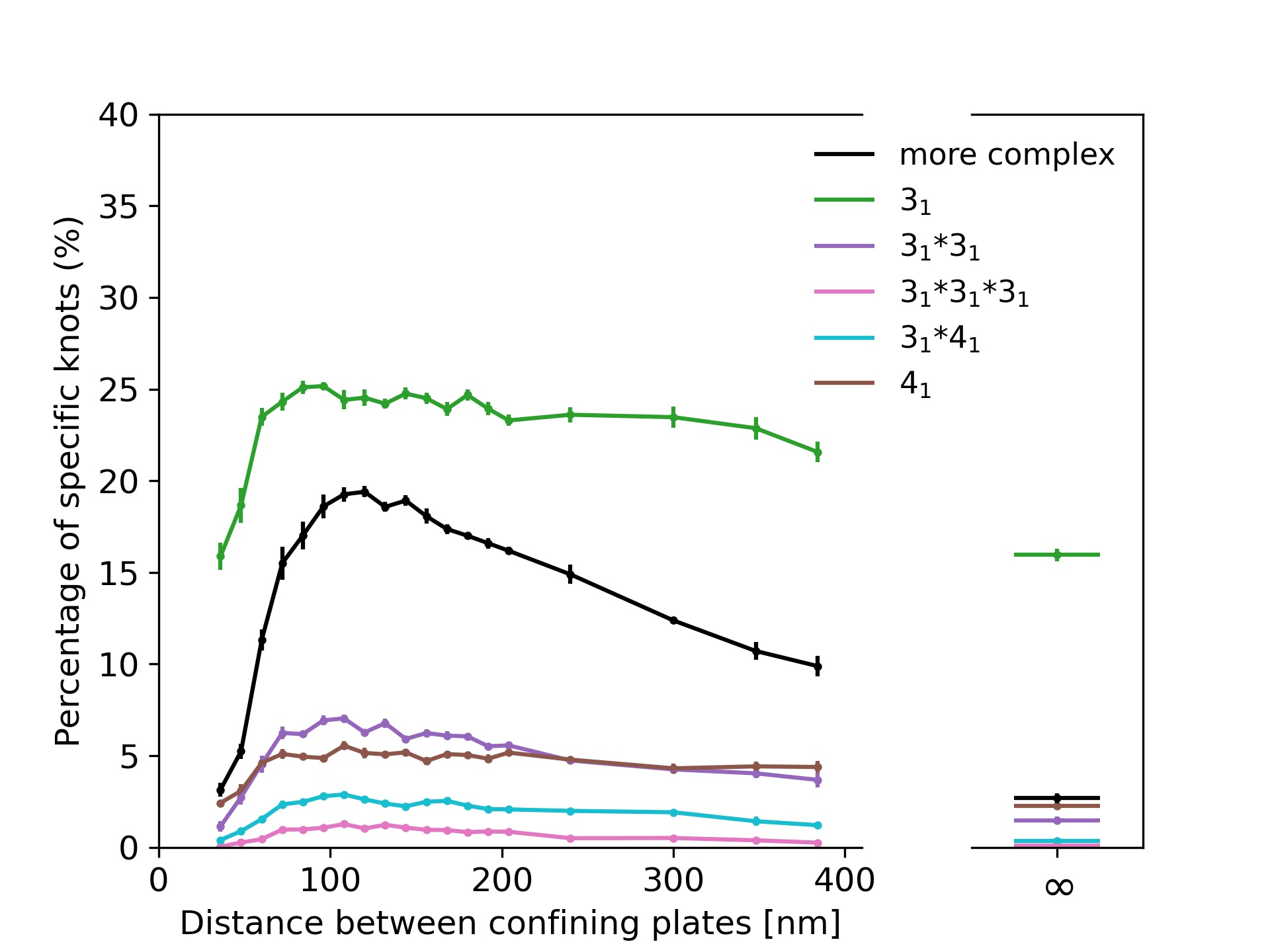}
		\label{fig:Subfigure3_2}
	\end{minipage}%

	\begin{minipage}[b]{0.6\textwidth}
		\includegraphics[width=\linewidth]{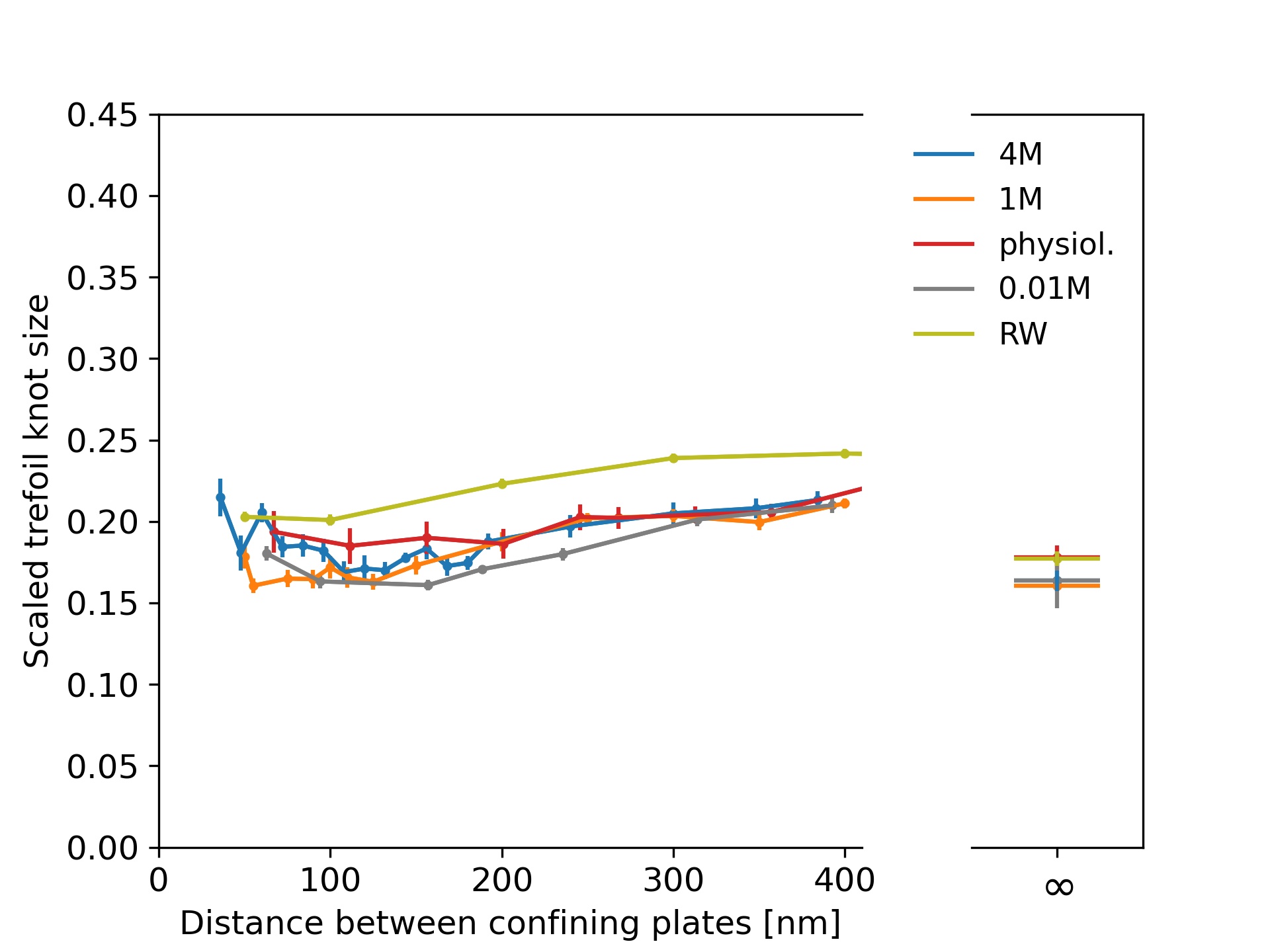}
		\label{fig:Subfigure3_3}
	\end{minipage}
	\caption{\textbf{(a)} Knotting probability of $\lambda$-phage DNA (48.5~kbp) as a function of plate distance at various ionic conditions in slit-pore confinement. 
		\textbf{(b)} Knot spectrum for a salt concentration of 4M. Complex knots refer to all other knots not listed here excluding the unknot. \textbf{(c)} Average sizes of trefoil knots.}
	\label{fig:fig3}
\end{figure}

In Fig.~\ref{fig:fig3}a we show results for $\lambda$-phage DNA (48,502 bp) confined between two plates to study the interplay of ionic conditions with confinement. As no experimental data is available, Fig.~\ref{fig:fig3}a only displays simulation results. The general shape of the curves follows results for shorter chains and high ionic conditions from \cite{Micheletti:SoftMatt:2012,Orlandini2013} and for rings in \cite{Dai_2012_ACS}: The knotting probability first increases with increasing plate distance, reaches a maximum at around 100 to 150~nm before falling off and approaching the value obtained for unconfined DNA. Here, we note again that knotting is suppressed substantially in low salt scenarios. 
For all salt concentrations, the number of knotted conformations in comparison to unconfined DNA is roughly increased by a factor of two at the maximum, and the position of the maximum shifts to lower plate distances with increasing salt concentrations as noted for closed rings in \cite{Dai_2012_ACS}.

Fig.~\ref{fig:fig3}b displays the knot spectrum as a function of plate distance for the 4M high salt scenario. 
While the amount of complex knots decreases (and unknots thus increase) for distances beyond the maximum, the composition of trefoil, figure-eight and composite variants of the two only varies slightly.

In Fig. \ref{fig:fig3}c, we plot the scaled trefoil knot length (which we define as the ratio of the contour length of the trefoil knot to the contour length of the whole chain). For all concentrations and plate distances, a trefoil knot roughly occupies one fifth of the chain and has a similar size as in the unconfined scenario. For a simple random walk, we roughly obtain the same result.

\section*{Discussion}
We investigate with numerical simulations the influence of ionic conditions on knotting of free DNA and DNA confined between two plates with a focus on long, experimentally relevant strands. From a technical point of view we test and confirm a coarse-grained bead-stick model by comparing simulations to recent nanopore experiments on DNA knotting. The model is not only susceptible to the influence of ionic conditions and reproduces the existing experimental knotting probabilities for unconfined DNA, but also resolves the structure of DNA below the persistence length. As such it is well-suited for the numerical description of recent \cite{Plesa2016,KumarSharma2019} and ongoing DNA experiments in the range of tens to hundreds of kilo base pairs and could be easily adapted for molecular dynamics simulations. Extensions which account for smaller, varying persistence lengths in small strands could be implemented as well to study structural properties of DNA at these scales \cite{Zoli_2018}.
At large length scales we observe a strong dependence on solvent conditions: While knotting can be abundant in a high salt scenario in which negative charges on DNA are completely screened, it becomes almost negligible in low salt conditions. Experiments on DNA dynamics \cite{Shusterman_2004} also imply that characteristic time scales involved in these transitions may well be below typical times required, e.g., for nanopore sequencing even though further studies on this issue are certainly warranted.
If this drastic change is confirmed experimentally in long strands, an adjustment of ionic conditions could indeed be used as a switch to effectively unknot DNA in scenarios where knots are undesired. Likewise, such experiments could further improve coarse-grained models by eliminating the need to assume effective excluded volume interactions, which could be fitted to knotting probabilities instead \cite{Rieger:PLoS:2016, Rieger_2018}.

\section*{Acknowledgement}
We are grateful to the Deutsche Forschungsgemeinschaft (DFG, German Research Foundation) for funding this research: Project number 233630050-TRR 146 and SFB 1551. The authors also acknowledge computing time granted on the supercomputer Mogon offered by the Johannes Gutenberg University Mainz (hpc.uni-mainz.de), which is a member of the AHRP (Alliance for High Performance Computing in Rhineland Palatinate,  www.ahrp.info) and the Gauss Alliance e.V. P.V. would also like to acknowledge helpful discussions with Eugene Kim.

\bibliographystyle{amsplain}
\bibliography{main}
\end{document}